\documentclass[a4paper,11pt]{article}
\pdfoutput=1 

\usepackage{jheppub} 
\usepackage{etoolbox}
    \makeatletter
    \patchcmd{\maketitle}{\@fpheader}{}{}{}
    \makeatother                     

\usepackage[T1]{fontenc} 
\usepackage{slashed}
\usepackage{epsfig,bbm}
\usepackage{amsmath}
\usepackage{slashed}
\usepackage{tikz}
\usepackage{bm}
\usepackage{amssymb}
\usepackage{mathtools}

\usepackage[compat=1.1.0]{tikz-feynman}

\newcommand{\be}{\begin{equation}}
\newcommand{\ee}{\end{equation}}
\newcommand{\bea}{\begin{eqnarray}}
\newcommand{\eea}{\end{eqnarray}}

\title{\boldmath  Self-resonant Dark Matter}

\preprint{}

\author[\dagger]{Seong-Sik Kim,}
\author[\dagger]{Hyun Min Lee,}
\author[\dagger,\ddagger]{and Bin Zhu}

\affiliation[\dagger]{Department of Physics, Chung-Ang University, Seoul 06974, Korea}
\affiliation[\ddagger]{School of Physics, Yantai University, Yantai 264005, China}

\emailAdd{sskim.working@gmail.com}
\emailAdd{hminlee@cau.ac.kr}
\emailAdd{zhubin@mail.nankai.edu.cn}

\abstract{
We present a novel mechanism for Sommerfeld enhancement for dark matter interactions without the need for light mediators. Considering a model for two-component dark matter with a triple coupling, we find that one of dark matter particles leads to an $u$-channel resonance in dark matter elastic scattering. From the sum of the $u$-channel ladder diagrams, we  obtain a Bethe-Salpeter equation with a delay term and identify the Sommerfeld factor from the elastic scattering of two dark matter components for the first time. We discuss the implications of our results for enhancing dark matter self-scattering and annihilation.   
}

\begin{document} 
\maketitle
\flushbottom

\section{Introduction}
Due to a variety of evidence for dark matter from astrophysics and cosmology, there have been a lot of particle physics models proposed for dark matter in the past. In particular, Weakly Interacting Massive Particles (WIMP) paradigm has drawn our community to search for dark matter accessible in particle physics experiments, with a motivation to solve some particle physics problems in the Standard Model such as the hierarchy problem. However, there has been no convincing evidence for direct detection of either dark matter or new particles beyond the Standard Model. Therefore, it is important to open a new avenue for pinning the properties of dark matter which are consistent with observations.   

Sommerfeld effects \cite{sommerfeld} have drawn a lot of attention due to their importance in performing the non-perturbative calculations for dark matter annihilation \cite{hisano,cirelli,russell,AH,pospelov,cassel,lengo,slatyer,feng,blum,kai}, dark matter self-scattering \cite{smallscale1,smallscale2,yu,smallscale3,slatyer2,zhang,kai2,kang,felix,sidm}, etc. Unlike the typical candidates for WIMP dark matter, Self-Interacting Dark Matter (SIDM) \cite{sidm0} has a large self-interaction to solve the small-scale problems at galaxies \cite{sidm} and also explains the diversity of galaxy rotation curves \cite{diversity}. Moreover, it is important to render the self-scattering of dark matter velocity-dependent to be consistent with the constraints from galaxy clusters \cite{smallscale3}. 

Sub-GeV scale dark matter can be good candidates for SIDM as for pions in QCD, whereas WIMP-like dark matter can be  self-interacting only with some enhancement mechanisms. 
In most cases considered in the literature until now, a sizable Sommerfeld enhancement for dark matter annihilations or a large non-perturbative effect for dark matter self-scattering were realized only due to the presence of light mediators that couple to dark matter.

In this letter, we propose a novel mechanism for obtaining the Sommerfeld enhancement for dark matter and making the dark matter self-scattering large without the need for light mediators.
We assume that dark matter is composed of at least two components with a triple interaction. In this case, we consider the elastic scattering between two dark matter components, for which one of dark matter particles is exchanged in the $u$-channel, and derive the Bethe-Salpeter equation for the system of two dark matter components in a form of delay differential equation.
From the resulting Schr\"odinger-like equation for two-component dark matter with a Yukawa-type interaction, we show that an effective light mediator from the off-shell dark matter particle gives rise to a sizable Sommerfeld factor at small velocities of dark matter.  

Our new enhancement mechanism for dark matter interactions does not require light mediators, so it is crucial to consider models with multi-component dark matter for such an unexpected new phenomenon.
We also discuss the velocity-dependent self-scattering for dark matter with  the $u$-channel resonance and comment on the implications of our results for dark matter annihilation processes such as $2\to 2$ and $3\to 2$ semi-annihilations.

\section{General idea}

We consider two dark matter components, $\phi_1$ and $\phi_2$, with masses satisfying $m_2<2m_1$ or $m_1<2m_2$ such that $\phi_1$ and $\phi_2$ are both stable and can be dark matter candidates, as far as there are no other decay channels for $\phi_1$ and $\phi_2$.  For a clear discussion below, assuming the mass ordering to be $m_2<2m_1$, we regard the elastic scattering between the two dark matter components, $\phi_1(q)\phi_2(p)\rightarrow \phi_1(q')\phi_2(p')$, as being exchanged by $\phi_1$ in the $u$-channel. 
A concrete model for two-component dark matter will be presented in the next section, but we first discuss the significance of dark matter particle exchanges in the $u$-channel in a model-independent way.

Assuming that the tree-level amplitude for the elastic scattering,  denoted as ${\widetilde \Gamma}(p,q;p',q')$, is exchanged by $\phi_1$ in the $u$-channel, and it is velocity-independent at the leading order in the expansion of dark matter momenta, we can take the tree-level scattering amplitude in the following form,
\bea
 {\widetilde \Gamma}(p,q;p',q') =\frac{4g^2 m^2_1}{|{\vec p}-{\vec q}'|^2+m^2_1-\omega^2}
\eea
where $g$ is a dimensionless coupling parameter and $\omega=p_0-q'_0$ is the energy exchange between two dark matter components.  
Therefore, in the non-relativistic limit for dark matter with  $m_2\neq m_1$, the the energy exchange $\omega$ becomes $\omega\approx m_2-m_1\neq 0$,  so the off-shell dark matter particle in the $u$-channel receives an effective squared mass, proportional to $m^2_1-\omega^2>0$ for $m_2< 2m_1$. As a result, for $m_2=2m_1$,  the off-shell dark matter particle  has an effectively vanishing masss, mediating a Coulomb-like long-range force between the two dark matter components.  Thus, we dub this possibility ``Self-resonant Dark Matter (SRDM)''.

Alternatively, we can take the same elastic scattering process to be exchanged by $\phi_2$ in the $u$-channel. In this case, the effective squared mass for the  off-shell dark matter particle is proportional to $m^2_2-\omega^2>0$ with $\omega\simeq m_2-m_1$, so it vanishes for $m_1=2m_2$.
Our general idea depends on the kinematics of dark matter particles, so it applies to general models for two component dark matter, independent of spins of dark matter.

\section{A model for two-component dark matter}

As illustration of the general idea, we consider a simple model for two-component scalar dark matter, composed of a complex scalar field $\phi_1$ and a real scalar field $\phi_2$, with a global or local $U(1)$ symmetry for $\phi_1$. The corresponding Lagrangian is in the following,
\bea
{\cal L}_s&=& |\partial_\mu\phi_1|^2 -m^2_1 |\phi_1|^2+\frac{1}{2}(\partial_\mu\phi_2)^2-\frac{1}{2} m^2_2 \phi^2_2 \nonumber \\
&&- 2g\,m_1 \phi_2 |\phi_1|^2.  \label{scalardm}
\eea
Then, the triple coupling between $\phi_1$ and $\phi_2$ contributes to the elastic scattering process, $\phi_1\phi_2\to \phi_1\phi_2$, and its complex conjugate process, at tree level, and the dark matter particle $\phi_1$ appears as an $u$-channel resonance. Extra couplings for dark matter such as $\phi^3_2$, $\phi_2^2|\phi_1|^2$, $|\phi_1|^4$ and $\phi^4_2$ can be included but they are not relevant for our later discussion on the $u$-channel resonance for  $\phi_1\phi_2\to \phi_1\phi_2$ or $\phi^*_1\phi_2\to \phi^*_1\phi_2$.  

The model suggested above is simple enough to capture the salient features of the $u$-channel resonance with a single dark matter particle, but it would be worthwhile to make a generalization of our model to dark matter models with more than two species and different spins. Concrete discussions on them will be work in progress and presented elsewhere \cite{zhu}. 

For the elastic scattering process with two-component scalar dark matter, $\phi_1(q)\phi_2(p)\rightarrow \phi_1(q')\phi_2(p')$,
the tree-level scattering amplitude ${\widetilde \Gamma}(p,q;p',q')$ becomes, in the non-relativistic limit,
\bea
 {\widetilde \Gamma}(p,q;p',q') &\approx& \frac{4g^2 m^2_1}{\Big(\sqrt{\frac{m_1}{m_2}} {\vec p}-\sqrt{\frac{m_2}{m_1}}{\vec q}'\Big)^2+m_2(2m_1-m_2)}  \nonumber \\
 &\equiv & U\bigg(\sqrt{\frac{m_1}{m_2}} {\vec p}-\sqrt{\frac{m_2}{m_1}}{\vec q}'\bigg). \label{tree4pt}
 \label{4point-tree}
\eea
Here, we remark that momentum-dependent terms in the energy transfer $\omega$ are included and the effective squared mass of the $u$-channel mediator in eq.~(\ref{tree4pt}) is semi-positive definite for $m_2\leq 2m_1$.  
For ${\vec p}={\vec q}'=0$, the above tree-level scattering amplitude diverges at $m_2=2m_1$, which we call {\it the $u$-channel resonance}. 
If $\phi_1$ is a stable dark matter, there is no decay width for $\phi_1$, so the above $u$-channel resonance cannot be regularized by a finite decay width, unlike the $s$-channel resonance in which case the mediator has a nonzero width in the region of a resonant enhancement.

 \begin{figure*}[tbp]
  \begin{center}
    \includegraphics[height=0.23\textwidth]{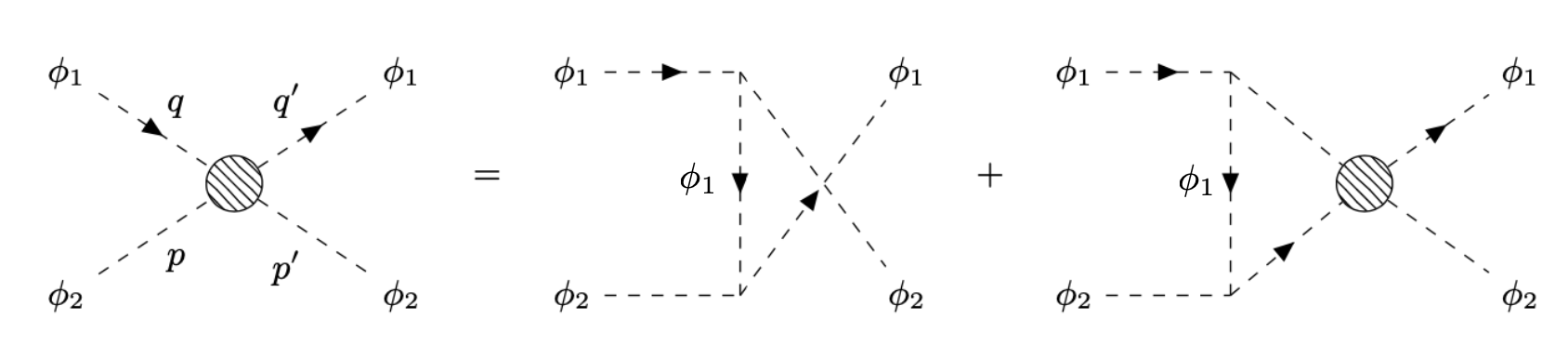}
  \end{center}
  \caption{Feynman diagrams for $\phi_1\phi_2\to \phi_1\phi_2$  at the non-perturbative level. }
  \label{nonpert}
\end{figure*}

We now consider the non-perturbative scattering amplitude for the elastic scattering process, $\phi_1(q)\phi_2(p)\rightarrow \phi_1(q')\phi_2(p')$. Thus, we need to include the ladder diagrams and obtain the non-perturbative four-point function $\Gamma(p,q;p',q')$ for the scattering process. 
To this purpose, we use the recursive relation for the  non-perturbative four-point function, as shown by the Feynman diagrams in Fig.~\ref{nonpert}, and define the following product for the four-point function,
\bea
\chi(p,q;p',q')\equiv G_2(p) G_1(q) \Gamma(p,q; p',q')\equiv \chi(p,q)
\eea
where  $G_{1,2}(p)$ are Feynman propagators for $\phi_{1,2}$, respectively.
Then, ignoring the perturbative contribution,  we obtain the approximate recursive relation for the non-perturbative four-point function as
\bea
i\chi(p,q)\approx -G_2(p)G_1(q) \int \frac{d^4 k}{(2\pi)^4} \,  {\widetilde \Gamma}(p,q;p+q-k,k)\, \chi(p+q-k,k). \label{BS0}
\eea

We now make a change of variables by
\bea
P=\frac{1}{2}(p+q),  \quad Q=\mu \Big(\frac{p}{m_2}-\frac{q}{m_1} \Big),
\eea
with $\mu=m_1 m_2/(m_1+m_2)$ being the reduced mass for  the $\phi_1-\phi_2$ system, and use the notation, 
$\chi(p,q)={\widetilde\chi}(P,Q)$.
Then, using the Bethe-Salpeter (BS) wave function \cite{SB,cassel} in momentum space,
${\widetilde\psi}_{BS}({\vec Q})= \int \frac{dQ_0}{2\pi} \,{\widetilde\chi}(P,Q)$,
we can rewrite the BS equation in  eq.~(\ref{BS0}) as
\bea
\left(\frac{{\vec Q}^2}{2\mu}-E \right){\widetilde\psi}_{BS}({\vec Q})=\frac{1}{4m_1m_2} \int \frac{d^3 k'}{(2\pi)^3} \, U\left(\left|\sqrt{\frac{m_1}{m_2}} {\vec Q}+\sqrt{\frac{m_2}{m_1}}{\vec k}'\right|\right)\, {\widetilde\psi}_{BS}({\vec k}')  \label{BS1}
\eea
where $P=\frac{1}{2}(m_1+m_2+E,0)$ and $Q=(Q_0,{\vec Q})$ in the center of mass coordinates with $E$ being the total kinetic energy.  
Then, making a Fourier transform for the BS wave function in position space, $\psi_{BS}({\vec x})$,
we get the BS equation in the following form,
\bea
-\frac{1}{2\mu}\,\nabla^2 \psi_{\rm BS}({\vec x}) + V({\vec x})\, \psi_{\rm BS}\Big(-\frac{m_2}{m_1}{\vec x}\Big)= E\psi_{\rm BS}({\vec x})  \label{BS2}
\eea
with
\bea
V({\vec x})=-\frac{\alpha}{r}\, e^{-M r}   \label{effpot}
\eea
where the effective mass is given by
\bea
M\equiv m_2\sqrt{2-\frac{m_2}{m_1}},  \label{effM}
\eea
$\alpha\equiv\frac{g^2}{4\pi}$ and $r=|{\vec x}|$. Therefore, the resulting potential is of the Yukawa type, with the effective mass $M$ in eq.~(\ref{effM}). For $m_2=2m_1$, the effective mediator mass vanishes, so a large Sommerfeld enhancement is anticipated as will be shown below. On the other hand, for almost degenerate masses, $m_2\simeq m_1$, the effective mass becomes $M\simeq m_1$, so there is no Sommerfeld enhancement without an extra light mediator \cite{slatyer2}.

\section{Sommerfeld factors for SRDM}

As the potential in the BS equation in eq.~(\ref{effpot}) is central, we can make a separation of variables  for the BS wavefunction as
\bea
\psi_{\rm BS}({\vec x})= R_{l}(r) Y_l^m(\theta,\phi). 
\eea
Then, we also get
\bea
\psi_{\rm BS}\Big(-\frac{m_2}{m_1}{\vec x}\Big)&=&  R_{l}\Big(\frac{m_2}{m_1}r\Big)Y_l^m(\pi-\theta,\phi+\pi) \nonumber \\
&=& (-1)^l R_{l}\Big(\frac{m_2}{m_1}r\Big) Y_l^m(\theta,\phi). 
\eea
As a result, after a change of variables, $x=\frac{1}{2}\mu \alpha r$, and a redefinition with
$R_{l}(x)= u_l(x)/x$, we can recast the BS equation (\ref{BS2}) to the following radial equation for $u_l$, 
\bea
\Bigg(\frac{d^2}{dx^2} -\frac{l(l+1)}{x^2}\Bigg) u_l(x)+\frac{4 e^{-c\,x}}{b\, x}  \, (-1)^l\, u_l(bx)+a^2\, u_l(x)=0 \label{xeq}
\eea
with
$a=\frac{2v_{\rm rel}}{\alpha}$,  $b=\frac{m_2}{m_1}$ and $c=\frac{2M}{\mu\alpha}$.
Here, $v_{\rm rel}$ is the relative velocity between $\phi_1$ and $\phi_2$. 
Therefore, we have got remarkable results in the radial equation. Namely, the $u$-channel interaction is attractive for even $l$, whereas it  becomes repulsive for odd $l$.

 \begin{figure}[tbp]
  \begin{center}
    \includegraphics[height=0.50\textwidth]{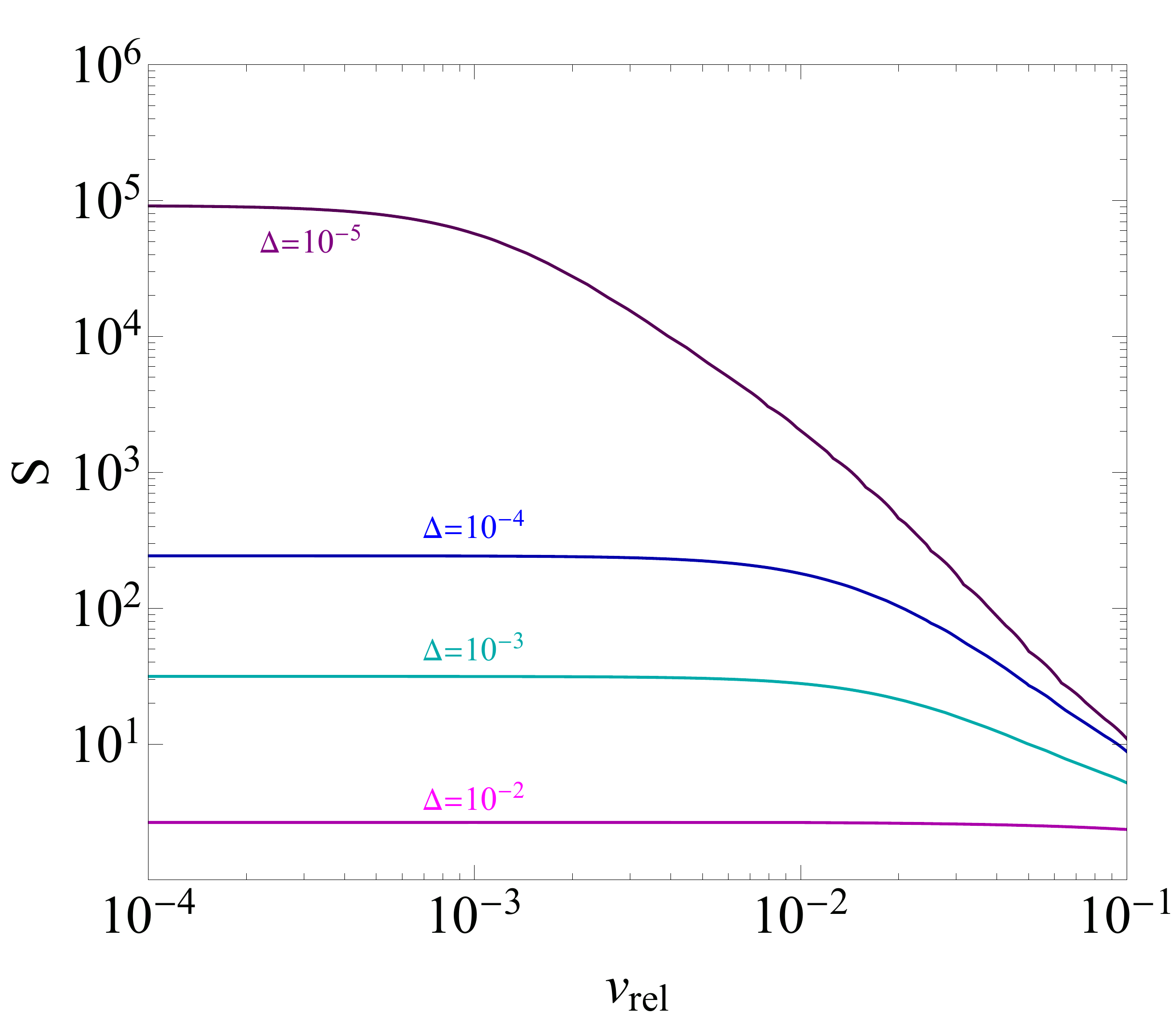}
  \end{center}
  \caption{Sommerfeld factor for $s$-wave elastic scattering, $\phi_1\phi_2\to \phi_1\phi_2$, as a function of the relative velocity $v_{\rm rel}$. We chose $\alpha=g^2/(4\pi)=0.1$ and $\Delta=10^{-2}, 10^{-3}, 10^{-4}, 10^{-5}$, for $m_2=2m_1(1-\Delta)$, in the lines from bottom to top.}
  \label{S-factor}
\end{figure}

Taking the Coulomb limit with $M=0$ (or $m_2=2m_1$) for $s$-wave with $l=0$, and making an change of variables, $x=e^{-\rho}$, we can recast eq.~(\ref{xeq}) into  a delay differential equations with constant shift, 
\bea
{\tilde u}^{\prime\prime}_0(\rho)+{\tilde u}^\prime_0(\rho)+2 e^{-\rho}\, {\tilde u}_0(\rho-\ln 2)+a^2e^{-2\rho}\, {\tilde u}_0(\rho)=0
\eea
with ${\tilde u}_0(\rho)=u_0(e^{-\rho})$ and ${\tilde u}^\prime_0$ being the derivative with respect to $\rho$, etc.  For $m_2< 2m_1$, the potential becomes a Yukawa-type and the shift in the above delay differential equation becomes $\ln b$.

We now impose the boundary conditions for the radial wave function to go to the plane-wave limit at $x=\infty$ and be finite at $x=0$, which correspond in the $\rho$ coordinate to
\bea
{\tilde u}_0(\rho)& \longrightarrow&\frac{1}{a}\, \sin(a\, e^{-\rho}+\delta_0), \qquad \rho \to -\infty,  \label{ubc1} \\
{\tilde u}_0(\rho)&\longrightarrow& A\, e^{-\rho}, \,\,\,\,\,\quad\qquad\qquad \rho\to +\infty \label{ubc2} 
\eea
where $k=\mu v_{\rm rel}$ and $\delta_0$ is the phase shift.
Then, after imposing the first boundary condition in eq.~(\ref{ubc1}) as a history function at $\rho=-\infty$ for the  delay differential equation, from the second boundary condition in eq.~(\ref{ubc2}), we can determine the phase shift $\delta_0$ as well as the Sommerfeld factor, $S_0=|R_0(0)|^2=A^2$. 

In Fig.~\ref{S-factor}, we show the Sommerfeld factor $S_0$ as a function of the relative velocity of dark matter, depending on $m_2=2m_1(1-\Delta)$ with $\Delta=10^{-2}, 10^{-3}, 10^{-4}, 10^{-5}$, in  pink, cyan, blue and purple lines, in order, and set  the triple coupling to $\alpha=g^2/(4\pi)=0.1$. Thus, we find that the closer the dark matter masses to the resonance condition, $m_2=2m_1$, the larger the Sommerfeld factor becomes at small velocities for dark matter. We remark that the Sommerfeld factor depends on $\Delta$ or the ratio of dark matter masses, but not on dark matter masses themselves, as can be seen from the model parameters, $a,b,c$, in eq.~(\ref{xeq}).

For nonzero $l$, we can also write the Schr\"odinger-like equation in eq.~(\ref{xeq}) with a Coulomb potential in the form of a delay differential equation with a delay constant $\ln b$, and impose similar boundary conditions at $\rho=\pm \infty$. In this case, however, only for even $l$, the effective potential is attractive, so there is a Sommerfeld enhancement factor.

\section{$2\to 2$ self-scattering for SRDM}

Taking $m_2\lesssim 2m_1$ for which the effective mass $M$ to be positive, 
we have the approximate Coulomb limit for $\phi_1\phi_2\to \phi_1\phi_2$ due to the $u$-channel resonance. 
In this case, since  the heavier particle $\phi_2$ is stable,  two components of dark matter, $\phi_1$ and $\phi_2$, coexist at present, so they undergo the elastic scattering.

 \begin{figure}[tbp]
  \begin{center}
    \includegraphics[height=0.50\textwidth]{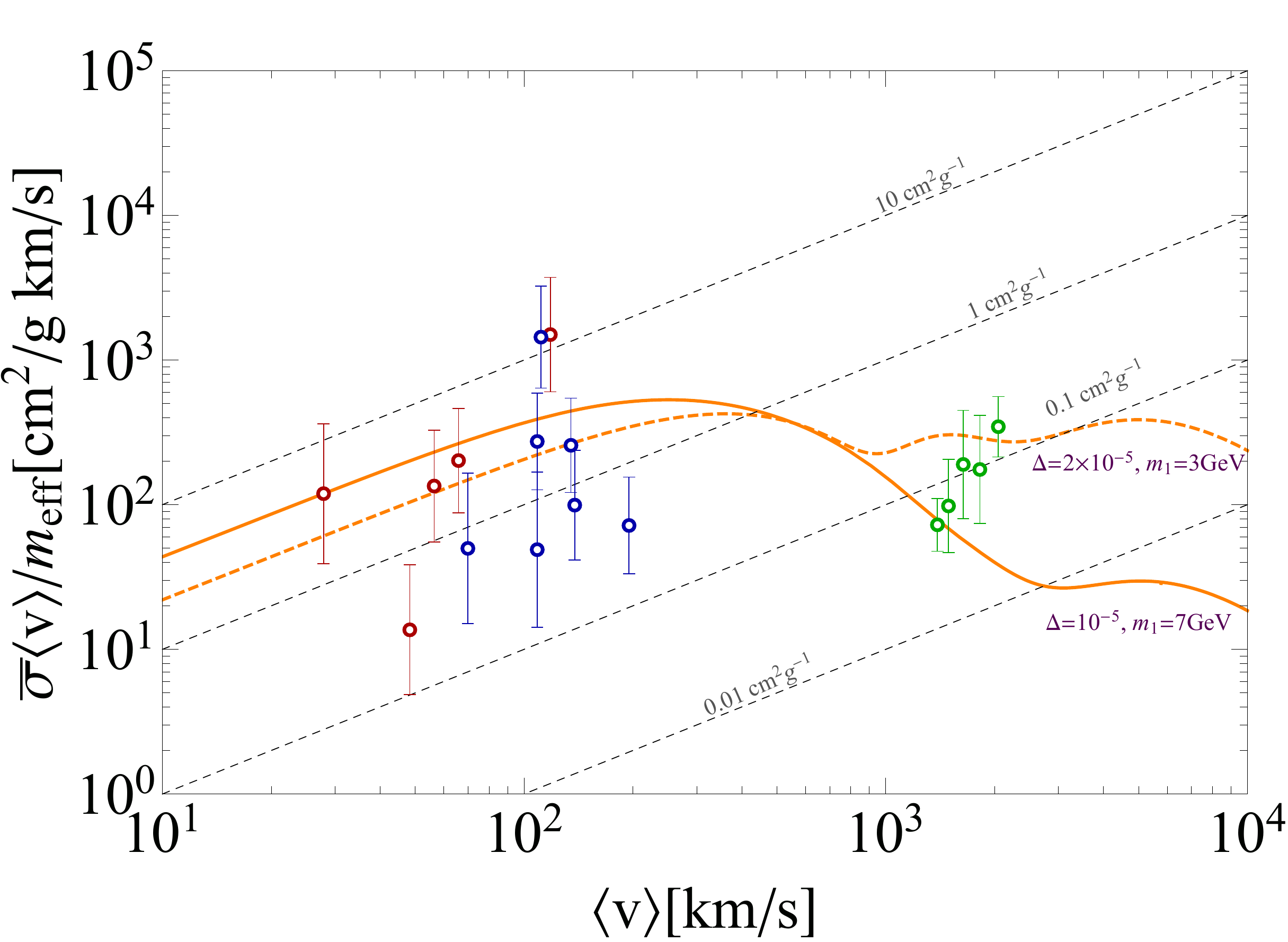}
  \end{center}
  \caption{Self-scattering cross section per dark matter mass for $s$-wave elastic scattering, $\phi_1\phi_2\to \phi_1\phi_2$, as a function of  $\langle v_{\rm rel}\rangle$. We chose $\Delta=10^{-5}, 2\times 10^{-5}$ and $m_1=7, 3\,{\rm GeV}$ in orange solid and dashed lines, respectively. We took $\alpha=g^2/(4\pi)=0.1$ and $m_{\rm eff}=2m_1(1+b)$. }
  \label{self}
\end{figure}

The elastic scattering cross-section for $\phi_1\phi_2\to \phi_1\phi_2$ can be enhanced by the non-perturbative effect coming from the effective light mediator in the $u$-channel.
For  $s$-wave dominance, the corresponding scattering amplitude is given by
$\sigma= \frac{4\pi}{k^2}\, \sin^2\delta_0$.
Therefore, we can achieve a large self-scattering cross section for dark matter up to unitarity bound, thanks to the $u$-channel resonance. In this case, the self-scattering cross section is enhanced at small velocities to solve the small-scale problems at galaxies and the diversity problem \cite{sidm}, while being consistent with the bounds from galaxy clusters \cite{smallscale3}.

In Fig.~\ref{self}, we depict the self-scattering cross section per effective dark matter mass as a function of the averaged relative velocity $\langle v_{\rm rel}\rangle$.  Here, we used the energy-transfer averaged self-scattering cross section with $\bar\sigma=\langle\sigma v_{\rm{rel}}^3\rangle/(24/\sqrt{\pi} v_0^3)$ with $v_0$ being the averaged velocity. We also took $m_{\rm eff}=2m_1(1+b)$ under the assumption that  two dark matter components are equally abundant at present, but the results would not change much as far as their abundances are comparable.  We set $\alpha=g^2/(4\pi)=0.1$ and chose the parameters for $m_2=2m_1(1-\Delta)$ with $\Delta=10^{-5}, 2\times 10^{-5}$ and $m_1=7, 3\,{\rm GeV}$ in orange solid and dashed lines, respectively. We indicated the required self-scattering cross sections from the data of rotational velocities at THINGS dwarf galaxies (red), LSB galaxies (blue), clusters (green) \cite{smallscale3}, and diagonal lines in gray show constant values of $\sigma/m$. In order to explain the rotational velocities of galaxies with self-scattering of dark matter with larger masses, we need to take smaller values of $\Delta$ for larger non-perturbative effects.

We note that there is another elastic scattering, $\phi_1\phi^{(*)}_1\rightarrow \phi_1\phi^{(*)}_1$ and its complex conjugate, which can have a similar resonance enhancement near $m_2\sim 2m_1$. 
However, the $s$-channel enhancement is less significant than $u$-channel resonance for $m_2\lesssim 2m_1$, because the center of mass energy for a pair of $\phi_1$ is always greater than $m_2$.
Therefore, the non-perturbative effects for  $\phi^{(*)}_1\phi_2\to \phi^{(*)}_1\phi_2$ give rise to dominant effects for rendering dark matter self-interacting. 

We remark that the non-perturbative effects in the $u$-channel resonance are also applicable for other dark matter candidates with different spins and they are responsible for enhancing dark matter annihilation processes, such as $2\to 2$ semi-annihilation, $\phi_1\phi_2\to \phi_1 X$, with $X$ being an extra mediator coupled to dark matter, and $3\to 2$ semi-annihilation, $\phi_1\phi_1\phi^*_1\to \phi_1 \phi_2$ \cite{vsimp}. We leave the related concrete discussion in another work \cite{zhu}.

\section{Conclusions}
We have proposed a new mechanism for Sommerfeld enhancement for dark matter without introducing a light mediator. We introduced a two-component scalar dark matter with a mass relation given by $m_2\lesssim 2m_1$. As a result, we showed that the lighter dark matter appears as an effectively massless resonance in the $u$-channel, namely, self-resonant dark matter, for $\phi_1\phi_2\to \phi_1\phi_2$ elastic scattering, and it gives rise to a Sommerfeld enhancement for dark matter. Due to a non-instantaneous interaction between two dark matter components, we showed that the resulting Bethe-Salpeter equation is cast into a form of delay differential equation, showing a non-linear enhancement of the Sommerfeld factor for a fixed plane-wave limit at large distances.

The proposed mechanism for Sommerfeld enhancement has immediate consequences for dark matter physics such as non-perturbative enhancements of dark matter self-scattering and annihilation. Furthermore, our results would shed light on the model building for dark matter with multiple components and make more precise calculations for related processes for dark matter possible.

\acknowledgments

We would like to thank Min-Seok Seo for discussion at the early stage of the project. 
The work is supported in part by Basic Science Research Program through the National Research Foundation of Korea (NRF) funded by the Ministry of Education, Science and Technology (NRF-2019R1A2C2003738). The work of BZ is supported partially by Korea Research Fellowship Program through the National Research Foundation of Korea (NRF) funded by the Ministry of Science and ICT (2019H1D3A1A01070937).
The work of SSK is supported by the Chung-Ang University Graduate Research Scholarship in 2020.

\appendix
\section{Derivation of the Bethe-Salpeter equation}

Here we provide the details for the derivation of the Bethe-Salpeter(BS) equation given in eq.~(\ref{BS2}) in the text, for a pair of dark matter components with different masses.
We focus on the elastic scattering process, $\phi_1(q)\phi_2(p)\rightarrow \phi_1(q')\phi_2(p')$, at non-perturbative level, and the same results are applicable to $\phi^*_1\phi_2\rightarrow \phi^*_1 \phi_2$. 

Summing up the ladder diagrams for $\phi_1\phi_2\rightarrow \phi_1 \phi_2$ with the triple interactions between dark matter components, we obtain the recursive relation for the non-perturbative four-point function $\Gamma(p,q;p',q')$ for the scattering process, as follows,
\bea
&&i\Gamma(p,q;p',q')=i {\tilde\Gamma}(p,q;p',q') \nonumber \\
&&\quad\qquad  -\int \frac{d^4k}{(2\pi)^4}\, {\widetilde \Gamma}(p,q;p+q-k,k) G_1(k)G_2(p+q-k) \Gamma(p+q-k,k;p',q')  \nonumber \\
&&\quad\approx -\int \frac{d^4k}{(2\pi)^4}\, {\widetilde \Gamma}(p,q;p+q-k,k) G_1(k)G_2(p+q-k) \Gamma(p+q-k,k;p',q') \label{a4point}
\eea
where $G_{1,2}(p)$ are Feynman propagators for $\phi_{1,2}$, given by
\bea
G_{1,2}(p)= \frac{i}{p^2-m^2_{1,2}},
\eea
and $k$ is the loop momentum, and we ignored the perturbative contributions in the approximation. 
Then, defining the following product of the  four-point function and the propagators,
\bea
\chi(p,q;p',q')\equiv G_2(p) G_1(q) \Gamma(p,q; p',q')\equiv \chi(p,q), 
\eea
and multiplying both sides of eq.~(\ref{a4point}) by $G_2(p) G_1(q)$,  we rewrite eq.~(\ref{a4point}) in momentum space,
\bea
i\chi(p,q)=-G_2(p)G_1(q) \int \frac{d^4 k}{(2\pi)^4} \,  {\widetilde \Gamma}(p,q;p+q-k,k)\, \chi(p+q-k,k). \label{aBS0}
\eea

We now make a change of variables by
\bea
P=\frac{1}{2}(p+q),  \quad Q=\mu \Big(\frac{p}{m_2}-\frac{q}{m_1} \Big),
\eea
with $\mu=m_1 m_2/(m_1+m_2)$ being the reduced mass for  the $\phi_1-\phi_2$ system,
or 
\bea
p=Q+\frac{2\mu}{m_1}\,P,  \quad q=-Q+\frac{2\mu}{m_2}\, P,
\eea
Here, $P,  Q$ are proportional to the velocity of the center of mass and the relative velocity, respectively. 
Then, noting
\bea
\chi(p,q)&=&{\widetilde\chi}(P,Q), \\
\chi(p+q-k,k)&=& {\widetilde\chi}\Big(P,\frac{2\mu}{m_2}P-k\Big).
\eea
where $\widetilde\chi$ is a function of $P$ and $Q$, eq.~(\ref{aBS0}) becomes
\bea
&&i{\widetilde\chi}(P,Q)=-G_2\Big(Q+\frac{2\mu}{m_1}\,P\Big) G_1\Big(-Q+\frac{2\mu}{m_2}\, P\Big) \nonumber \\
&&\times \int \frac{d^4 k}{(2\pi)^4} \,  {\widetilde \Gamma}(p,q;p+q-k,k) \, {\widetilde\chi}\Big(P,\frac{2\mu}{m_2}P-k\Big) \nonumber \\
&=&-G_2\Big(Q+\frac{2\mu}{m_1}\,P\Big) G_1\Big(-Q+\frac{2\mu}{m_2}\, P\Big) \nonumber \\
 &\times&\int \frac{d^4 k'}{(2\pi)^4} \,  {\widetilde \Gamma}(p,q;p+q-k,k)\Big|_{k=-k'+\frac{2\mu}{m_2}P} \, {\widetilde\chi}(P,k')  \label{aBS00}
\eea
where we made a shift in loop momentum  by $k'=-k+\frac{2\mu}{m_2}P$ in the second equality, and the tree-level four-point function given in eq.~(\ref{4point-tree}) becomes
\bea
 {\widetilde \Gamma}(p,q;p+q-k,k)\Big|_{k=-k'+\frac{2\mu}{m_2}P}&=&  \frac{4g^2 m^2_1}{\Big(\sqrt{\frac{m_1}{m_2}} {\vec p}-\sqrt{\frac{m_2}{m_1}}{\vec k}\Big)^2+m_2(2m_1-m_2)} \Bigg|_{k=-k'+\frac{2\mu}{m_2}P}\nonumber \\
 &=&  \frac{4g^2 m^2_1}{\Big(\sqrt{\frac{m_1}{m_2}} {\vec Q}+\sqrt{\frac{m_2}{m_1}}{\vec k}'\Big)^2+m_2(2m_1-m_2)}  \nonumber \\
 &\equiv& U\left(\left|\sqrt{\frac{m_1}{m_2}} {\vec Q}+\sqrt{\frac{m_2}{m_1}}{\vec k}'\right|\right).
\eea

Using the BS wave function in momentum space,
\bea
{\widetilde\psi}_{BS}({\vec Q})= \int \frac{dQ_0}{2\pi} \,{\widetilde\chi}(P,Q),
\eea
and multiplying eq.~(\ref{aBS00}) by $\int dQ_0/(2\pi)$, we get
\bea
i{\widetilde\psi}_{BS}({\vec Q})&=& - \int \frac{dQ_0}{2\pi} G_2\Big(Q+\frac{2\mu}{m_1}\,P\Big) G_1\Big(-Q+\frac{2\mu}{m_2}\, P\Big) \times\nonumber \\
&&\quad\times  \int \frac{d^3 k'}{(2\pi)^3} \,  U\left(\left|\sqrt{\frac{m_1}{m_2}} {\vec Q}+\sqrt{\frac{m_2}{m_1}}{\vec k}'\right|\right)\, {\widetilde\psi}_{BS}({\vec k}') \nonumber \\
&=&-\frac{i}{4m_1 m_2 \Big(E-\frac{{\vec Q}^2}{2\mu}\Big)}\, \int \frac{d^3 k'}{(2\pi)^3} \, U\left(\left|\sqrt{\frac{m_1}{m_2}} {\vec Q}+\sqrt{\frac{m_2}{m_1}}{\vec k}'\right|\right)\, {\widetilde\psi}_{BS}({\vec k}'). 
\eea
Here, we chose the center of mass coordinates for which $P=\frac{1}{2}(m_1+m_2+E,0)$ and $Q=(Q_0,{\vec Q})$, with $E$ being the total kinetic energy.  
Therefore, we get
\bea
\left(\frac{{\vec Q}^2}{2\mu}-E \right){\widetilde\psi}_{BS}({\vec Q})=- \int \frac{d^3 k'}{(2\pi)^3} \, {\widetilde V}_1\left(\left|\sqrt{\frac{m_1}{m_2}} {\vec Q}+\sqrt{\frac{m_2}{m_1}}{\vec k}'\right|\right)\, {\widetilde\psi}_{BS}({\vec k}')  \label{BS1a}
\eea
with ${\widetilde V}_1=-U/(4m_1m_2)$.
Then, using the BS wave function in position space,
\bea
\psi_{BS}({\vec x})=\int\frac{d^3{\vec Q}}{(2\pi)^3}\, e^{i{\vec Q}\cdot {\vec x}} {\widetilde\psi}_{BS}({\vec Q}),
\eea
 and multiplying both sides of the above equation by $\int \frac{d^3 {\vec Q}}{(2\pi)^3}\, e^{i{\vec Q}\cdot {\vec x}}$, we get
\bea
&&\left(-\frac{1}{2\mu}\nabla^2-E \right)\psi_{BS}({\vec x}) \nonumber \\
&=& -\int \frac{d^3 {\vec Q}}{(2\pi)^3}\, e^{i{\vec Q}\cdot {\vec x}} \int \frac{d^3 k'}{(2\pi)^3} \, {\widetilde V}_1\left(\left|\sqrt{\frac{m_1}{m_2}} {\vec Q}+\sqrt{\frac{m_2}{m_1}}{\vec k}'\right|\right)\, {\widetilde\psi}_{BS}({\vec k}') \nonumber \\
&=& -\int \frac{d^3 {\vec Q}}{(2\pi)^3}\, e^{i{\vec Q}\cdot {\vec x}} \int \frac{d^3 k'}{(2\pi)^3}\int d^3{\vec x}'\, {\rm exp}\left[-i\Big(\sqrt{\frac{m_1}{m_2}} {\vec Q}+\sqrt{\frac{m_2}{m_1}}{\vec k}'\Big)\cdot {\vec x}' \right]  V_1({\vec x}') \, {\widetilde\psi}_{BS}({\vec k}') \nonumber \\
&=&-\int d^3{\vec x}'\, \delta^3\Big({\vec x}-\sqrt{\frac{m_1}{m_2}} {\vec x}'\Big) \int  \frac{d^3 k'}{(2\pi)^3} {\rm exp}\left(-i\sqrt{\frac{m_2}{m_1}}{\vec k}'\cdot {\vec x}'\right)  V_1({\vec x}') {\widetilde\psi}_{BS}({\vec k}') \nonumber \\
&=&-\Big( \frac{m_2}{m_1}\Big)^{3/2}V_1\Big(\sqrt{\frac{m_2}{m_1}}{\vec x} \Big)  \int  \frac{d^3 k'}{(2\pi)^3}\,  {\rm exp}\left(-i\,\frac{m_2}{m_1}{\vec k}'\cdot {\vec x}\right)  {\widetilde\psi}_{BS}({\vec k}') \nonumber \\
&=& -\Big( \frac{m_2}{m_1}\Big)^{3/2}V_1\Big(\sqrt{\frac{m_2}{m_1}}{\vec x} \Big)  {\psi}_{BS}\Big(-\frac{m_2}{m_1}{\vec x}\Big) \equiv -V({\vec x} )  {\psi}_{BS}\Big(-\frac{m_2}{m_1}{\vec x}\Big).
\eea
Therefore, we obtain the BS equation in the following form,
\bea
-\frac{1}{2\mu}\,\nabla^2 \psi_{\rm BS}({\vec x}) + V({\vec x})\, \psi_{\rm BS}\Big(-\frac{m_2}{m_1}{\vec x}\Big)= E\psi_{\rm BS}({\vec x})  \label{aBS2}
\eea
with
\bea
V({\vec x})&=&\Big( \frac{m_2}{m_1}\Big)^{3/2}V_1\Big(\sqrt{\frac{m_2}{m_1}}{\vec x} \Big)=-\frac{1}{4m_1m_2}\,\Big( \frac{m_2}{m_1}\Big)^{3/2} U\Big(\sqrt{\frac{m_2}{m_1}}{\vec x} \Big)  \nonumber \\
&=&-\frac{1}{4m_1m_2}\Big( \frac{m_2}{m_1}\Big)^{3/2} \int \frac{d^3{\vec q}}{(2\pi)^3}\, {\rm exp}\left(i \sqrt{\frac{m_2}{m_1}}{\vec q}\cdot {\vec x}\right)\,\frac{4g^2 m^2_1 }{{\vec q}^2+m_2 (2m_1-m_2)} \nonumber \\
&=&-\frac{\alpha}{r}\, e^{-M r}  
\eea
where $\alpha\equiv\frac{g^2}{4\pi}$ and the effective mass for the $u$-channel propagator is given by
\bea
M\equiv m_2\sqrt{2-\frac{m_2}{m_1}}. 
\eea
Here, use is made of $ \int \frac{d^3{\vec q}}{(2\pi)^3}\, e^{i{\vec q}\cdot {\vec x}}/({\vec q}^2+m^2)=e^{-mr}/(4\pi r)$. 
The above discussion completes the derivation of eq.~(\ref{BS2}), which is the key equation for our discussion on the Sommerfeld enhancement and the non-perturbative self-scattering for dark matter in our model.


\begin{thebibliography}{999}



\bibitem{sommerfeld}
A. Sommerfeld, 
Ann. Phys. 403 (1931) 257.


\bibitem{hisano}
J.~Hisano, S.~Matsumoto, M.~M.~Nojiri and O.~Saito,
Phys. Rev. D \textbf{71} (2005), 063528
doi:10.1103/PhysRevD.71.063528 
[arXiv:hep-ph/0412403 [hep-ph]].

\bibitem{cirelli}
M.~Cirelli, A.~Strumia and M.~Tamburini,
Nucl. Phys. B \textbf{787} (2007), 152-175
doi:10.1016/j.nuclphysb.2007.07.023
[arXiv:0706.4071 [hep-ph]].


\bibitem{russell}
J.~D.~March-Russell and S.~M.~West,
Phys. Lett. B \textbf{676} (2009), 133-139
doi:10.1016/j.physletb.2009.04.010
[arXiv:0812.0559 [astro-ph]].

  
\bibitem{AH}
N.~Arkani-Hamed, D.~P.~Finkbeiner, T.~R.~Slatyer and N.~Weiner,
Phys. Rev. D \textbf{79} (2009), 015014
doi:10.1103/PhysRevD.79.015014
[arXiv:0810.0713 [hep-ph]].


\bibitem{pospelov}
M.~Pospelov and A.~Ritz,
Phys. Lett. B \textbf{671} (2009), 391-397
doi:10.1016/j.physletb.2008.12.012
[arXiv:0810.1502 [hep-ph]].


\bibitem{cassel}
S.~Cassel,
J. Phys. G \textbf{37} (2010), 105009
doi:10.1088/0954-3899/37/10/105009
[arXiv:0903.5307 [hep-ph]].
  
  
\bibitem{lengo}
R.~Iengo,
JHEP \textbf{05} (2009), 024
doi:10.1088/1126-6708/2009/05/024
[arXiv:0902.0688 [hep-ph]].
  


\bibitem{slatyer}   
T.~R.~Slatyer,
JCAP \textbf{02} (2010), 028
doi:10.1088/1475-7516/2010/02/028
[arXiv:0910.5713 [hep-ph]].
      
   
\bibitem{feng}
J.~L.~Feng, M.~Kaplinghat and H.~B.~Yu,
Phys. Rev. D \textbf{82} (2010), 083525
doi:10.1103/PhysRevD.82.083525
[arXiv:1005.4678 [hep-ph]].
      
      

\bibitem{blum}
K.~Blum, R.~Sato and T.~R.~Slatyer,
JCAP \textbf{06} (2016), 021
doi:10.1088/1475-7516/2016/06/021
[arXiv:1603.01383 [hep-ph]].

\bibitem{kai}
T.~Bringmann, F.~Kahlhoefer, K.~Schmidt-Hoberg and P.~Walia,
Phys. Rev. Lett. \textbf{118} (2017) no.14, 141802
doi:10.1103/PhysRevLett.118.141802
[arXiv:1612.00845 [hep-ph]].


\bibitem{smallscale1}
J.~L.~Feng, M.~Kaplinghat and H.~B.~Yu,
Phys. Rev. Lett. \textbf{104} (2010), 151301
doi:10.1103/PhysRevLett.104.151301
[arXiv:0911.0422 [hep-ph]].



\bibitem{smallscale2}
S.~Tulin, H.~B.~Yu and K.~M.~Zurek,
Phys. Rev. Lett. \textbf{110} (2013) no.11, 111301
doi:10.1103/PhysRevLett.110.111301
[arXiv:1210.0900 [hep-ph]].
  
  
  
\bibitem{yu}  
S.~Tulin, H.~B.~Yu and K.~M.~Zurek,
Phys. Rev. D \textbf{87} (2013) no.11, 115007
doi:10.1103/PhysRevD.87.115007
[arXiv:1302.3898 [hep-ph]].




\bibitem{smallscale3}
M.~Kaplinghat, S.~Tulin and H.~B.~Yu,
Phys. Rev. Lett. \textbf{116} (2016) no.4, 041302
doi:10.1103/PhysRevLett.116.041302
[arXiv:1508.03339 [astro-ph.CO]].



\bibitem{slatyer2}
K.~Schutz and T.~R.~Slatyer,
JCAP \textbf{01} (2015), 021
doi:10.1088/1475-7516/2015/01/021
[arXiv:1409.2867 [hep-ph]].


\bibitem{zhang}
Y.~Zhang,
Phys. Dark Univ. \textbf{15} (2017), 82-89
doi:10.1016/j.dark.2016.12.003
[arXiv:1611.03492 [hep-ph]].


\bibitem{kai2}
F.~Kahlhoefer, K.~Schmidt-Hoberg and S.~Wild,
JCAP \textbf{08} (2017), 003
doi:10.1088/1475-7516/2017/08/003
[arXiv:1704.02149 [hep-ph]].


\bibitem{kang}
Y.~J.~Kang and H.~M.~Lee,
J. Phys. G \textbf{48} (2021) no.4, 045002
doi:10.1088/1361-6471/abe529
[arXiv:2003.09290 [hep-ph]].


\bibitem{felix}
B.~Colquhoun, S.~Heeba, F.~Kahlhoefer, L.~Sagunski and S.~Tulin,
Phys. Rev. D \textbf{103} (2021) no.3, 035006
doi:10.1103/PhysRevD.103.035006
[arXiv:2011.04679 [hep-ph]].



\bibitem{sidm}
S.~Tulin and H.~B.~Yu,
Phys. Rept. \textbf{730} (2018), 1-57
doi:10.1016/j.physrep.2017.11.004
[arXiv:1705.02358 [hep-ph]].



\bibitem{sidm0}    
D.~N.~Spergel and P.~J.~Steinhardt,
Phys. Rev. Lett. \textbf{84} (2000), 3760-3763
doi:10.1103/PhysRevLett.84.3760
[arXiv:astro-ph/9909386 [astro-ph]].
    

\bibitem{diversity}    
A.~Kamada, M.~Kaplinghat, A.~B.~Pace and H.~B.~Yu,
Phys. Rev. Lett. \textbf{119} (2017) no.11, 111102
doi:10.1103/PhysRevLett.119.111102
[arXiv:1611.02716 [astro-ph.GA]];
M.~Kaplinghat, T.~Ren and H.~B.~Yu,
JCAP \textbf{06} (2020), 027
doi:10.1088/1475-7516/2020/06/027
[arXiv:1911.00544 [astro-ph.GA]].



\bibitem{zhu}
S.-S. Kim, H. M. Lee and B. Zhu, To appear.



\bibitem{SB}
E.~E.~Salpeter and H.~A.~Bethe,
Phys. Rev. \textbf{84} (1951), 1232-1242
doi:10.1103/PhysRev.84.1232




\bibitem{vsimp}
S.~M.~Choi, Y.~Hochberg, E.~Kuflik, H.~M.~Lee, Y.~Mambrini, H.~Murayama and M.~Pierre,
JHEP \textbf{10} (2017), 162
doi:10.1007/JHEP10(2017)162
[arXiv:1707.01434 [hep-ph]];
S.~M.~Choi, H.~M.~Lee, Y.~Mambrini and M.~Pierre,
JHEP \textbf{07} (2019), 049
doi:10.1007/JHEP07(2019)049
[arXiv:1904.04109 [hep-ph]].




 

\end{thebibliography}
\end{document}